\documentclass[12pt]{iopart}
\begin{document}
\title
{How to simulate a universal quantum computer using negative probabilities}

\author{Holger F. Hofmann}

\address{Graduate School of Advanced Sciences of Matter, Hiroshima University,
Kagamiyama 1-3-1, Higashi Hiroshima 739-8530, Japan}
\ead{hofmann@hiroshima-u.ac.jp}
\begin{abstract}
The concept of negative probabilities can be used to decompose
the interaction of two qubits mediated by a quantum controlled-NOT 
into three operations that require only classical interactions 
(that is, local operations and classical communication)
between the qubits. For a single gate, the probabilities of the three
operations are $1$, $1$, and $-~1$. This decomposition can be applied 
in a probabilistic simulation of quantum computation by randomly choosing
one of the three operations for each gate and assigning a negative 
statistical weight to the outcomes of sequences with an odd number of
negative probability operations. The maximal exponential speed-up of 
a quantum computer can then be evaluated in terms of the increase in 
the number of sequences needed to simulate a single operation of the quantum circuit.
\end{abstract}

\maketitle


\maketitle

\section{Introduction}
The observation that quantum systems can under some circumstances 
outperform comparable classical systems is a central
motivation of quantum information research. In one of the earliest
proposals of quantum computation, Richard Feynman pointed out that 
the difficulty of simulating quantum systems on a classical computer
was evidence of the superior efficiency of a quantum computer
\cite{Fey82}. In the
same presentation, he also describes an attempt to simulate quantum
statistics by decomposing the density matrix into probabilities.
Since this decomposition results in negative probabilities, the 
conclusion is that a classical simulation of quantum probabilities
is not possible. 
Nevertheless, negative probabilities can be a useful
tool in the ``resolution'' of quantum paradoxes such as the violation of
Bell`s inequalities \cite{Han96,Cer00,Hof01} and the observation 
of measurement 
results outside the normal range in weak measurements 
\cite{Hof00,Mit07,Sov07,Mir07}.
Since such paradoxes appear to be closely related to the efficiency
of quantum computation, it may be worthwhile to update Feynman's
negative probability approach to quantum computation by describing 
the operation of a universal quantum computer in terms of negative 
probabilities. 

Specifically, negative probabilities can be used to decompose 
the entangling multi-qubit gates of a universal 
quantum computer into statistical mixtures of non-entangling 
local operations. 
These non-entangling local operations can be simulated 
efficiently by a classical probabilistic computation that only
needs to keep track of the local qubit states, e.g. by representing 
them as classical spins, as suggested in the context of NMR quantum 
computation \cite{Sch99}. It is thus possible to represent the
quantum statistics of the computation entirely in terms of the classical
statistics of an analogous spin system, simply by including a single
additional marker bit that distinguishes negative from positive 
probability contributions. 
By choosing a minimally negative decomposition of the
entangling gate operation, it is possible to design a 
classical probabilistic simulation of the quantum computation that
not only produces the correct output statistics, but also 
allows a step-by-step analysis of the computation. 
The effects of the non-local quantum
coherence expressed by entangled states can thus be represented in 
terms of negative probabilities of entirely local states. 
Since negative probabilities add up just like positive ones, 
the correct output probabilities of the quantum formalism can be 
obtained from the output frequencies of the classical simulation 
by simply assigning a negative statistical weight to sequences with an 
odd number of negative probability operations, providing a recipe for
stochastic simulations that emphasizes the similarity of the quantum
formalism with classical statistics.

In the following, it is shown
that a quantum controlled-NOT gate can be expressed as a 
statistical mixture of three local operations with probabilities of 
$1$, $1$, and $-~1$, which is the minimal negativity for this
entangling gate. The gate operation can then
be simulated classically
by attributing a negative statistical weight to those outcomes 
that were obtained from operations with negative probability. Since the 
quantum controlled-NOT gate is universal in the sense that any 
quantum computation
can be constructed using only quantum controlled-NOT gates and local
operations \cite{DiV95}, this negative probability decomposition can be 
applied to obtain a classical probabilistic simulation of any multi-qubit 
quantum operation. In this simulation, each circuit with $N$ two qubit 
gate operations is described by a set of $3^N$ sequences of local operations 
with positive and
negative probabilities. As a result, the statistical relevance of each
individual outcome is reduced by a factor of $1/3^N$, and the number
of classical runs needed to simulate a single run of the quantum 
circuit increases exponentially with the number of two qubit gates. 
The direct comparison of the universal quantum computer using quantum controlled-NOT gates and its corresponding classical probabilistic
simulation therefore indicates that in principle, an exponential
speed-up of up to $3^N$ may be achieved by the use of entangling
gate operations.

\section{Local decomposition of a single entangling gate}
The starting point for any negative probability decomposition of quantum
operations is the process matrix representation, which corresponds to 
the density matrix representation for quantum states. 
The elements $\chi_{ij}$ of the process matrix of a quantum operation 
on a $d$-dimensional Hilbert space are defined using a basis set of
$d^2$ orthonormal operators $\hat{A}_i$. The effect of the operation 
$E$ on an arbitrary input density operator $\hat{\rho}$ is then given by
\begin{equation}
E(\hat{\rho})=\sum_{i,j=1}^d \chi_{ij} \hat{A}_i \hat{\rho} 
\hat{A}_j^\dagger.
\end{equation}
In the case of separate systems, the operators $\hat{A}_i$ are usually
defined by products of local basis operators. In such a local operator
basis, a completely diagonal process matrix represents a mixture of 
correlated local operations with no entanglement capability. However, 
entangling operations have coherences between their local components that
cannot be represented by positive mixtures of local products. Any
decomposition into a weighted sum of local operations will therefore 
include some negative weights. 

In the case of two qubit operations, a convenient set of basis
operators is given by the two qubit products of the Pauli operators 
$X$, $Y$, $Z$, and the identity $I$ \cite{NCtext}. 
By themselves, these operators describe $\pi$-rotations around the
corresponding axes of the Bloch vectors representing the qubits.
All other operations are described by coherent superpositions
of these operators. In particular, the quantum controlled-NOT 
operation is given by the coherent superposition
\begin{equation}
\hat{U}_{\mbox{\small CNOT}} = \frac{1}{2}\left(I \otimes I 
+ Z \otimes I + I \otimes X - Z \otimes X \right).
\end{equation}
The process matrix of the quantum controlled-NOT therefore includes 
maximal coherences between all
four basis operations. Taken separately, these coherences
can also be obtained from local operations on the two qubits, but the
combination of all of the coherences results in a gate with maximal
entanglement capability. 

As shown previously elsewhere \cite{Hof05a}, the process matrix of 
the quantum controlled-NOT gate can be decomposed into a sum of three local 
operations reproducing the coherences and a negative dephasing
term that effectively restores the full coherence of the 
original quantum gate. It is then possible to identify specific sets 
of coherences with directly observable local operation, providing 
an experimental
criterion for quantum parallelism \cite{Bao07}. In the present context
however, the goal is to minimize the negativity of the decomposition.
This can be achieved by combining the negative dephasing operation
with one of the positive operations. The process matrix is then 
decomposed into only three local components: two positive ones with the 
same coherence as the quantum controlled-NOT, and one negative one with 
the opposite coherence,
\begin{equation}
\label{eq:decomp}
E_{\mbox{\small CNOT}}(\hat{\rho}) = 
L_1(\hat{\rho})+L_2(\hat{\rho})-\bar{L}_3(\hat{\rho}).
\end{equation}
For reasons of symmetry, it is convenient to choose the coherences
$\chi_{II,ZI}$ ($\chi_{II,IX}$) and $\chi_{IX,ZX}$ ($\chi_{ZI,ZX}$)
to define the positive operation $L_1$ ($L_2$), and the coherences 
$\chi_{II,ZX}$ and $\chi_{IX,ZI}$ to define the negative 
operation $\bar{L}_3$. 
The first set of coherences is described by the local operation
\begin{equation}
\begin{array}{c}
L_1(\hat{\rho})=\hat{M}_{Z0}\; \hat{\rho}\; \hat{M}_{Z0}
+\hat{M}_{Z1}\; \hat{\rho}\; \hat{M}_{Z1}
\\[0.2cm]
\hat{M}_{Z0}=\frac{1}{2}(I+Z) \otimes I; \hspace{0.3cm}
\hat{M}_{Z1}=\frac{1}{2}(I-Z) \otimes X.
\end{array}
\end{equation}
This operation describes a local measurement of $Z$ on qubit 1,
followed by a conditional rotation $X$ on qubit 2 if the result
was $-1$, which corresponds to a logical $1$ of the control qubit. 
It is thus a local implementation of the controlled-NOT operation
in the computational basis.
Similarly, the second set of coherences is described by the local
operation
\begin{equation}
\begin{array}{c}
L_2(\hat{\rho})=\hat{M}_{X0}\; \hat{\rho}\; \hat{M}_{X0}
+\hat{M}_{X1}\; \hat{\rho}\; \hat{M}_{X1}
\\[0.2cm]
\hat{M}_{X0}=\frac{1}{2}I \otimes (I+X); \hspace{0.3cm}
\hat{M}_{X1}=\frac{1}{2}Z \otimes (I-X).
\end{array}
\end{equation}
This operation describes a local measurement of $X$ on qubit 2, 
followed by a conditional rotation $Z$ on qubit 1 if the 
result was $-~1$. It is a local implementation of the reverse
controlled-NOT operation observed in the $X$ basis, which is
complementary to the operation in the computational basis
\cite{Hof05b,Oka05}. 
Finally, the third set of coherences is described by the negative 
operation
\begin{equation}
\begin{array}{c}
\bar{L}_3(\hat{\rho})=
\frac{1}{2}
\hat{U}_{a}\; \hat{\rho}\; \hat{U}_{a}^\dagger
+ \frac{1}{2}
\hat{U}_{b}\; \hat{\rho}\; \hat{U}_{b}^\dagger
\\[0.2cm]
\hat{U}_{a}=\frac{1}{2}(I+iZ) \otimes (I-iX); \hspace{0.3cm}
\hat{U}_{b}=\hat{U}_{a}^{-1}.
\end{array}
\end{equation}
This operation describes a correlated pair of $\pi/2$ rotations
around the $Z$ and $X$ axes of qubit 1 and 2, respectively. 
Since this is the operation with negative probability, the
correlation between the rotations is opposite to the one
that can be observed in the actual operation of a quantum 
controlled-NOT \cite{Bao07}.
 
For the following analysis, it is essential that the decomposition
given above has the lowest possible negativity possible for an
input state independent decomposition of the quantum controlled-NOT 
gate. Specifically, it needs to be shown that the remaining negative
probability of $-1$ is the minimal negativity necessary to explain the
entanglement capability of the gate. Since the 
quantum controlled-NOT can generate a maximally entangled state
from local inputs, this problem is equivalent to showing that the 
minimal negativity of a local decomposition for a maximally entangled 
two qubit state is $-1$. Since it is well known that the maximal
overlap between a local stateand a maximally entangled state of two 
qubits is $F=1/2$, the overlap of the maximally entangled state
with a normalized mixture of local states with positive probabilities
of $1+n$ and negative probabilities of $-~n$ is limited to 
$F\leq (1+n)/2$. Hence, a negative probability of at least $-~1$
is necessary for a local decomposition of the maximally entangled state,
and likewise for the quantum controlled-NOT
or any other maximally entangling two qubit gate.

The discussion above proves that the negative probability
in eq.(\ref{eq:decomp}) is the minimal negativity necessary for a
representation of the entanglement capability of the gate. The negativity
of the decomposition is therefore a direct measure of the non-local
content of the gate operation \cite{Eis00,Col01}. Further
reductions of negativity are only possible if some assumptions are made
about the possible input states. That is, the decomposition should be 
optimal for any quantum circuit designed to efficiently process 
completely arbitrary input states. In conventional quantum circuits,
more efficient simulations are possible because the input states 
are either well defined, or limited to eigenstates of the 
computational basis. Thus, conventional quantum circuits are usually designed to process only classical information, restricting their 
operation to a tiny segment of the available Hilbert space. The 
simulation proposed here has the advantage that it can be applied 
without further analysis of these restrictions imposed on a specific
circuit. It thus applies even to a universal quantum computer able to
process quantum information directly, without state preparation and
measurement.

The decomposition given above can be interpreted as a negative probability 
mixture of three local operations with probabilities of $p(L_1)=p(L_2)=1$ 
and $p(\bar{L}_3)=-1$ that reproduces the non-local unitary operation of 
the quantum controlled-NOT gate. It is therefore possible to obtain the
correct output statistics of the quantum gate by adding the output 
probabilities of the local operations $L_1$ and $L_2$ and subtracting the 
output probabilities of the local operation $\bar{L}_3$. 
The operation of a non-local quantum gate can then be simulated by
peforming only local operations and classical communication between
the qubits. 

\section{Probabilities for sequences of gate operations}
To decompose an arbitrarily complex quantum circuit, all we need to 
do is to evaluate the total probability of a sequence of $N$ 
gate operations. 
Since the gate operations are linear, and since the output density
matrix can be written as a linear combination of the outputs from the
three local operations, it is possible to apply the conventional rules 
of Bayesian statistics. The statistical weight of a sequence $i$ of 
the local operations
$L_1$, $L_2$ and $\bar{L}_3$ is therefore equal to the product of the
statistical weights of each operation. Since the statistical weights
of $L_1$, $L_2$ and $\bar{L}_3$ are $+1$, $+1$, and $-1$, 
the probability $p(i)$ of a 
particular sequence $i$ of local operations is given by
\begin{equation}
p(i) = \left\{
\begin{array}{ccl}
+1 && \mbox{for even numbers of $\bar{L}_3$}
\\
-1 && \mbox{for odd numbers of $\bar{L}_3$}
\end{array}
\right.
\end{equation}
In total, there are $3^N$ possible sequences $i$. Specifically,
there are $(3^N-1)/2$ negative probability sequences
with an odd number of $\bar{L}_3$ operations, and $(3^N+1)/2$ positive
probability sequences with an even number of $\bar{L}_3$ operations,
for a total probability of one.

In a classical simulation, both positive and negative sequences must
be performed with equal (naturally positive) frequency.
The probabilities of the quantum process $p_{\mbox{\small quant.}}$
are therefore related to the classical simulation probabilities 
$p_{\mbox{\small pos.}}$ and $p_{\mbox{\small neg.}}$ of the positive and negative 
sequences by
\begin{equation}
\label{eq:qclass}
p_{\mbox{\small quant.}}=3^N (p_{\mbox{\small pos.}}-p_{\mbox{\small neg.}}).
\end{equation}
Here, the amplification factor of $3^N$ expresses the different
normalizations of the classical probabilities and the quantum
probabilities. Specifically, the classical simulation necessarily
replaces the negative frequencies $f_{\mbox{\small quant.}}=k p_i$
associated with negative probabilities with positive frequencies
$f_{\mbox{\small neg.}}=k |p_i|$. As a result, the ratio of the
total number of trials needed for the classical simulation to the
$k$ trials used in the quantum process becomes
\begin{equation}
\label{eq:ratio}
\frac{\sum f_{\mbox{\small pos.}}+\sum f_{\mbox{\small neg.}}}
{\sum f_{\mbox{\small quant.}}}=
\frac{\sum_i |p(i)|}{\sum_i p(i)}=3^N.
\end{equation}
The possibility of achieving an exponential speed-up by quantum
computation is therefore directly observable as an 
amplification factor of $3^N$ that relates the probability differences
in the classical simulation to the probability differences observed in
the actual quantum operation. 
It may be worth noting that this result is closely related 
to the exponential 
decay of the signal predicted by classical models of NMR quantum 
computations, as reported in \cite{Sch99}. In fact,
(\ref{eq:qclass}) and (\ref{eq:ratio}) can be interpreted as 
representations of the minimal signal decrease caused by
the classical simulation of the entangling gates, providing a
quantitative expression for the conjecture at the end of \cite{Sch99} 
that ``an ultimate signal decrease is the consequence of any attempt 
to describe entangling unitaries classically''.

The correct output of the quantum circuit can be written as a statistical
mixture of the outputs of the $3^N$ local operations, with the appropriate
statistical weights of $p(i)=\pm 1$. The probability $p(m)$ of obtaining
a specific measurement result $m$ in the output can thus be written
according to standard Bayesian probability theory as
\begin{equation}
\label{eq:condprob}
p(m) = \sum_i p(m|i) p(i),
\end{equation}
where $p(m|i)$ is the conditional output probability of sequence $i$
determined from the mixture of local output states obtained by applying
sequence $i$. (\ref{eq:condprob}) represents the quantum analog of 
classical causality, showing that the introduction of negative 
probabilities permits a detailed analysis of quantum operations in terms
of well separated, non-interfering classical sequences of events.
Since the introduction of negative probabilities $p(i)$ is the minimal negative probability necessary to
obtain a local description of the entangling gates, this should be the
closest possible analogy between quantum and classical processes that
works for any combination of local gates and quantum controlled-NOT 
operations.

\section{Simulation of an entanglement paradox}
One of the main merits of the negative probability decomposition is the
representation of entanglement effects in terms of negative probability 
mixtures of local alternatives.  
To see how this decomposition ``resolves'' the
paradoxical aspects of entanglement, it may be instructive to 
take a closer look at the example of a specific quantum circuit 
generating a maximally entangled state. 

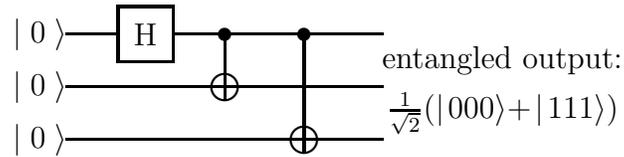
\begin{figure}[t]
\begin{center}
\begin{picture}(240,70)
\thicklines
\put(10,40){\makebox(20,20)
{$\mid 0 \; \rangle$}}
\put(10,20){\makebox(20,20)
{$\mid 0 \; \rangle$}}
\put(10,0){\makebox(20,20)
{$\mid 0 \; \rangle$}}

\put(50,40){\framebox(20,20){H}}
\put(30,50){\line(1,0){20}}
\put(70,50){\line(1,0){80}}
\put(30,30){\line(1,0){120}}
\put(30,10){\line(1,0){120}}
\put(90,50){\circle*{5}}
\put(90,50){\line(0,-1){25}}
\put(90,30){\circle{10}}
\put(120,50){\circle*{5}}
\put(120,50){\line(0,-1){45}}
\put(120,10){\circle{10}}

\put(150,30){\makebox(90,20)
{entangled output:}}
\put(150,10){\makebox(90,20)
{$\frac{1}{\sqrt{2}}(\mid \! 000 \rangle +\! \mid \! 111 \rangle)$}}

\end{picture}
\end{center}
\caption{\label{circuit}
Quantum logic circuit generating a maximally entangled three qubit 
GHZ state.}
\end{figure}

One of the most simple 
cases is the circuit shown in figure \ref{circuit}, which generates 
the three qubit Greenberger-Horne-Zeilinger (GHZ) state
$(\mid~0~0~0~\rangle + \mid~1~1~1~\rangle)/\sqrt{2}$ from a 
non-entangled product of $|0\rangle$ states by a sequence of 
one Hadamard gate on qubit one and two quantum controlled-NOT gates
that change the states of qubit 2 and qubit 3 according to the state 
of the control qubit 1. The three qubit GHZ state has the paradoxical property that the product of its $X$ values is always $+1$, but the 
product of any one $X$ value and the two remaining $Y$ values
are always $-1$. Since only three of these four properties can be
true for any simultaneous assignments of $X$ and $Y$ values to the
three qubits, the four probabilities of $1$ observed in the
GHZ state output are a striking proof of the impossibility of
local hidden variable models \cite{Gre89,Mer90,Ber99}. 
It should therefore be interesting to see how the gate 
operations generate the four correlations.


\begin{table}[th]
\caption{\label{result}
Contributions of the nine sequences of local operations
of the circuit in figure \ref{circuit} to the 
output probabilities of the four correlations
of the GHZ paradox. 
}
\begin{indented}
\item[]\begin{tabular}{cccccc}
\br
\multicolumn{2}{c}{Sequence} & \multicolumn{4}{c}{$p(m|i)$ for $m$ given by}
\\
& \hspace{0.8cm} 
& \hspace{0.2cm} $XXX$ \hspace{0.2cm}
& \hspace{0.2cm} $XYY$ \hspace{0.2cm}
& \hspace{0.2cm} $YXY$ \hspace{0.2cm}
& \hspace{0.2cm} $YYX$ \hspace{0.2cm}
\\[-0.1cm]
 $i$ & $p(i)$ & $=+1$ & $=-1$ & $=-1$ & $=-1$
\\ \mr
$L_1-L_1$ & 1 & 1/2 & 1/2 & 1/2 & 1/2
\\
$L_1-L_2$ & 1 & 1/2 & 1/2 & 1/2 & 1/2
\\
$L_1-\bar{L}_3$ & -1 & 1/2 & 1/2 & 1/2 & 1/2
\\ \mr
$L_2-L_1$ & 1 & 1/2 & 1/2 & 1/2 & 1/2
\\
$L_2-L_2$ & 1 & 1 & 1/2 & 1/2 & 1/2
\\
$L_2-\bar{L}_3$ & -1 & 1/2 & 1/2 & 0 & 1/2
\\ \mr
$\bar{L}_3-L_1$ & -1 & 1/2 & 1/2 & 1/2 & 1/2
\\
$\bar{L}_3-L_2$ & -1 & 1/2 & 1/2 & 1/2 & 0
\\
$\bar{L}_3-\bar{L}_3$ & 1 & 1/2 & 1 & 1/2 & 1/2
\\ \br
\end{tabular}
\end{indented}
\end{table}

We can find this out by 
decomposing the two gate operations into nine sequences of local 
operations, identifying the conditional probabilities $p(XXX=+1|i)$,
$p(XYY=-1|i)$, $p(YXY=-1|i)$, and $p(YYX=-1|i)$. The results are shown in
table \ref{result}. Specifically, each correlation can be traced to a
different sequence of local operations, with $L_2-L_2$ generating 
$XXX=+1$ and $L_3-L_3$ generating $XYY=-1$, while the negative probability
operations $L_2-\bar{L}_3$ and $\bar{L}_3-L_2$ generate $YXY=+1$ and 
$YYX=+1$, reducing the total probabilities of these outputs to zero
and leaving a probability of 1 for the opposite results of 
$YXY=-1$ and $YYX=-1$.
In all four cases, the remaining eight sequences of local operations
result in probabilities of $1/2$. Thus the GHZ paradox is ``resolved'' 
by separating the sequences that generate the four correlations.
This is possible because the negative probabilities in
(\ref{eq:condprob}) allow conditional
probabilities of $p(m|i) \neq 1$ even when the total probability is 
$p(m)=1$.

\section{Quantum parallelism and uncertainty}
The example in the previous section illustrates how negative probabilities
can restore locality to the description of quantum processes. 
Without changing the mathematical structure of the formalism, it is thus
possible to represent the non-local coherences of the Hilbert space
formalism as non-interfering negative probability mixtures of local 
alternatives. The advantage of this approach is that it establishes a
very close analogy between classical probabilistic computation and 
the use of entanglement in quantum computation. This may be especially
useful for the interaction between experiment and theory, since
the experimental verification of quantum processes is usually
based on local measurement statistics. Hence the effects of entanglement
are obtained by combining the correlations observed in separate 
measurements, based on the notion that the outcomes observed separately 
on identically prepared systems all represent equally valid 
features of the actual quantum process occurring in parallel. 
The simulation of quantum processes by negative probabilities corresponds
to this empirical notion of quantum parallelism, providing a description 
of quantum processes that is closer to the experimentally 
accessible evidence than the Hilbert space formalism \cite{Hof05a,Bao07}.

Since it is obvious that actual measurement outcomes can never have 
negative probabilities, it may be appropriate to reflect a bit on 
the justifications for the use of negative probabilities in quantum 
mechanics. In the example of the three qubit GHZ paradox, experiments
are limited to measuring either $X$ or $Y$. As a result of this
uncertainty limit on local quantum measurements, it is not
possible to construct an actual experiment where the negative probabilities
of table \ref{result} would result in an impossible prediction.
Thus the uncertainty principle ``covers up'' the negative probabilities
and the assignment of negative values to joint probabilities of $X$ and $Y$ 
can be consistent with the experimental evidence.
One of the consequences of this possibility is that it invalidates the
claim of Einstein, Podolsky, and Rosen \cite{EPR} that an element of 
reality must be attributed to measurement outcomes that that can be 
predicted with 100\% certainty. Instead, probabilities
of 100 \% can still be conditional, since they may arise from a 
cancellation of negative and positive joint probabilities for 
alternative results. It is therefore possible to resolve entanglement
paradoxes if one is willing to give up the notion of a non-empirical
reality beyond the uncertainty limit. 

In the case of the GHZ state generation, it is clear that the output 
measurements cannot distinguish between the nine sequences of 
operations given in table \ref{result}, making the
details of the quantum process experimentally inaccessible.
The assignment of negative probabilities $p(i)$ is then
consistent with all possible measurement results. 
In the classical simulation, the output probabilities are 
determined by separating positive and negative sequences and
assigning attenuated positive probabilities of 
$p_{\mbox{\small pos.}}(i)=p_{\mbox{\small neg.}}(i)=1/9$. 
It is then possible to
trace the probability of $p_{\mbox{\small quant.}}
(XXX=+1)=1$ to the difference between
$p_{\mbox{\small pos.}}(XXX=+1)=3/9$ and $p_{\mbox{\small neg.}}(XXX=+1)=2/9$,
amplified by a factor of 9 according to (\ref{eq:qclass}).
Thus, the classical simulation makes the details of the process
accessible at a cost of reduced probabilities and hence reduced
certainty about the results. 
Effectively, there appears to be a fundamental trade-off between 
the uncertainty limited access to details of the quantum process
and the enhanced precision of the observable outcomes. This
trade-off can be expressed in the ``currency'' 
of potentially negative probabilities. 
In the context of the present work, this indicates that 
uncertainty about the actual sequence of logical operations is 
the price to be paid for the possibility of an exponential 
speed-up in quantum computation. 

\section{On the universality of negative probability simulations}
The essential feature of the simulation presented in this paper is
that it can be applied to arbitrary networks of controlled-NOT gates
and local unitaries. Since this is a universal set of gates, any
arbitrary quantum process can be simulated in this manner. 
It is certainly possible to find more efficient simulations for 
specific processes and algorithms, but such simulations would 
depend on specific features of the processes (e.g. a restriction
of gate operations such as the one considered in the Gottesmann-Knill
theorem \cite{NCtext}). However, no conceivable quantum process can 
exceed the speed-up given by the probability amplification
of $3^N$, since there always exists a corresponding classical process
that reproduces $p_{\mbox{\small quant.}}$ in terms of 
$p_{\mbox{\small pos.}}$
and $p_{\mbox{\small neg.}}$ according to (\ref{eq:qclass}). 
Thus the probability amplification of $3^N$
provides an upper limit for the possible computational speed-up
of universal quantum computers.

In addition to this quantitative limit, the universal 
correspondence between classical probabilistic computation and 
quantum computation established by the use of negative probabilities 
may also provide a key to the microscopic analysis of quantum
effects beyond the uncertainty limit. 
In particular, the negative probability analysis describes
conditional probabilities that could be tested experimentally,
either by interrupting a statistical sample of the computations by projective 
measurements \cite{Mor06}, or by using weak measurements 
with negligible back-action \cite{Aha88,Rit91,Res04}. 
In the case of weak measurements, the correspondence of the non-classical 
features to negative probabilities is already well established (see e. g.
\cite{Hof00,Mir07}), and the possible implications for quantum computation 
have recently been discussed in \cite{Mit07}. Moreover, it has been shown
both theoretically and experimentally that weak measurements can be
performed using a quantum controlled-NOT
gate to implement the measurement interaction \cite{Pry05,Ral06}.
It is therefore possible to describe the weak measurement as a part of
the quantum circuit, 
providing a ``resolution'' of the paradox 
of post-selected weak values outside 
the range of possible eigenvalues in terms of the quantum
parallelism of the two qubit gate. 

Finally, it may be worth considering the possible extensions of 
the present work to other entangling interactions.
In principle, the present approach is not limited to the 
quantum controlled-NOT gate, and its extension to other
universal gates such as the recently realized quantum 
Toffoli gate \cite{Lan08} may provide further insights into 
the general relation between entanglement capability and 
computational speed-up. On a more fundamental level, it may 
also be interesting
to consider a direct simulation of quantum interactions using
a negative probability decomposition of the interaction mediated 
by a Hamiltonian into conditional local operations \cite{Hof06}. 
This approach might establish a general connection between 
computational speed-up and the interaction dynamics of quantum systems.

\section{Conclusions}
It has been shown that a universal quantum computer can be simulated
by a closely related classical probabilistic computation, where the
effects of entanglement are simulated by assigning negative statistical
weights to a well defined part of the outcomes.
Since each two qubit gate is simulated by a selection of one of three
local operations, the exponential speed-up of a quantum computation 
involving $N$ two qubit gates can be described by the factor of $3^N$,
representing the number of possible sequences of local operations
needed to simulate a single quantum operation of the circuit. 
This statistical expression of exponential speed-up represents an
upper limit for any speed-up achieved in quantum computation. 

In addition to providing a fairly simple and compact recipe for
a classical simulation of any arbitrary quantum process, the negative
probability decomposition described in this paper can also be used 
to analyze general non-classical features of quantum processes.
As illustrated by the example given above, it is then possible to
explain not only the possibility of computational speed-up, but also the
paradoxical aspects of entanglement and weak measurement in a 
single unifying framework. The simulation of quantum computation
by negative probabilities may therefore be the key to a
more intuitive and consistent understanding of quantum systems
in general.

\section*{Acknowledgements}
I would like to thank A. Hosoya for his questions on
negative probabilities that motivated me to write this paper.
Part of this work has been supported by the Grant-in-Aid program of the
Japanese Society for the Promotion of Science and by the JST-CREST
project on quantum information processing.

\end{document}